\documentclass[pdflatex,sn-mathphys]{sn-jnl}

\jyear{2021}
\theoremstyle{thmstyleone}

\theoremstyle{thmstyletwo}%

\theoremstyle{thmstylethree}%
\usepackage{amsbsy}
\usepackage{wrapfig}
\usepackage{framed}
\normalbaroutside

\begin{document}

\title[Bayesian method for estimating the Hurst exponent]{Optimizing a Bayesian method for estimating the Hurst exponent in behavioral sciences}

\author*[1]{\fnm{Madhur} \sur{Mangalam}}\email{mmangalam@unomaha.edu}

\author[1]{\fnm{Taylor} \sur{Wilson}}

\author[2]{\fnm{Joel} \sur{Sommerfeld}}

\author*[1]{\fnm{Aaron D.} \sur{Likens}}\email{alikens@unomaha.edu}

\affil[1]{\orgdiv{Division of Biomechanics and Research Development, Department of Biomechanics, and Center for Research in Human Movement Variability}, \orgname{University of Nebraska at Omaha}, \orgaddress{\street{University Dr S}, \city{Omaha}, \postcode{68182}, \state{NE}, \country{USA}}}

\abstract{The Bayesian Hurst-Kolmogorov (HK) method estimates the Hurst exponent of a time series more accurately than the age-old detrended fluctuation analysis (DFA), especially when the time series is short. However, this advantage comes at the cost of computation time. The computation time increases exponentially with $N$, easily exceeding several hours for $N = 1024$, limiting the utility of the HK method in real-time paradigms, such as biofeedback and brain-computer interfaces. To address this issue, we have provided data on the estimation accuracy of $H$ for synthetic time series as a function of \textit{a priori} known values of $H$, the time series length, and the simulated sample size from the posterior distribution---a critical step in the Bayesian estimation method. The simulated sample from the posterior distribution as small as $n = 25$ suffices to estimate $H$ with reasonable accuracy for a time series as short as $256$ measurements. Using a larger simulated sample from the posterior distribution---i.e., $n > 50$---provides only marginal gain in accuracy, which might not be worth trading off with computational efficiency. We suggest balancing the simulated sample size from the posterior distribution of $H$ with the computational resources available to the user, preferring a minimum of $n = 50$ and opting for larger sample sizes based on time and resource constraints.}


\maketitle

\section*{Introduction}

A robust measure of the strength of long-range correlations in time series is the Hurst exponent, $H$, named by Mandelbrot \cite{mandelbrot1969computer} in honor of pioneering work by Edwin Hurst in hydrology \cite{hurst1951long}. In the parlance of linear statistics, $H$ quantifies how the measurements' $SD$-like variations grow across progressively longer timescales, indicating how the correlation among sequential measurements might decay across longer separations in time. $H$ describes a single fractal-scaling estimate of power-law decay in autocorrelation $\rho$ for lag $k$ as $\rho_{k} = |k + 1|^{2H} -2|k|^{2H} + |k - 1|^{2H}$, for which $H$ reveals the degree of persistence ($0.5 < H < 1.0$; i.e., large values are typically followed by large values) or anti-persistence ($0 < H < 0.5$; i.e., small values typically follow large values and vice versa).

\vspace{6 pt}

$H$ has become a central inferential statistic in diverse fields, including meteorology \cite{efstathiou2010altitude,ivanova1999application,tatli2020long}, economics \cite{alvarez2008short,grau2000empirical,ivanov2004common,liu1997correlations,liu1999statistical}, ethology \cite{alados2000fractal,bee2001individual}, bioinformatics \cite{buldyrev1998analysis,mantegna1994linguistic,peng1993finite}, and physiology \cite{castiglioni2019fast,goldberger2002fractal,hardstone2012detrended,peng1993long}. In behavioral sciences, successful examples of inferences made using the $H$ statistic include interpretations about feedforward and forward processes in postural control \cite{delignieres2011transition,duarte2008complexity,lin2008reliability}, system-wide coordination \cite{chen1997long,diniz2011contemporary}, cognition \cite{allegrini2009spontaneous,gilden1995noise,kello2010scaling,stephen2008strong,van2003self}, and perception-action \cite{mangalam2019fractal,mangalam2020bodywide,mangalam2020global}, among countless others. $H$ has also proved to be an effective measure differentiating among adults with healthy and pathological cardiovascular functioning \cite{ashkenazy1999discrimination,ho1997predicting,peng1995fractal}, as well as movement systems \cite{bartsch2007fluctuation,hausdorff1997altered,hausdorff2001human,hausdorff2007gait,herman2005gait,kobsar2014evaluation}. $H$ is also becoming a statistical benchmark for developing rehabilitative interventions \cite{mangalam2022leveraging,raffalt2021temporal,raffalt2023stride} and quantifying the effectiveness of those interventions \cite{kaipust2013gait,marmelat2020fractal,vaz2020gait}.

\vspace{6 pt}

The most common method of estimating $H$ is detrended fluctuation analysis (DFA) \cite{peng1994mosaic,peng1995quantification}. DFA’s ability to assess the strength of long-range correlations embedded in time series that seem non-stationary and to prevent the false detection of long-range correlations that are a byproduct of non-stationarity make it superior to many other methods. Numerical analysis has shown that DFA confers several advantages when the data trend's functional form is not known \textit{a priori} \cite{bashan2008comparison,grech2005statistical}. Nonetheless, DFA has several shortcomings which none of the existing alternatives overcome \cite{almurad2016evenly,delignieres2006fractal,dlask2019hurst,katsev2003hurst,marmelat2019fractal,ravi2020assessing,roume2019biases,schaefer2014comparative,yuan2018unbiased}. For instance, DFA does not accurately assess the strength of long-range correlations when the time series is brief \cite{dlask2019hurst,katsev2003hurst,schaefer2014comparative}, producing a positive bias in its central tendency in addition to a large dispersion \cite{almurad2016evenly,delignieres2006fractal,marmelat2019fractal,ravi2020assessing,roume2019biases,yuan2018unbiased}. DFA requires time series consisting of at least $500$ measurements to accurately estimate $H$, severely limiting its application under time constraints or when collecting longer time series is not practical, such as in pathological populations who cannot participate in a study for an extended time due to fatigue \cite{marmelat2019fractal}. Furthermore, DFA is precariously sensitive to the time series length, typically overestimating $H$, a trend present in long time series but exaggerated when used with brief time series\cite{delignieres2006fractal,stroe2009estimating}.

\vspace{6 pt}

An alternative approach to estimating $H$---not well-known in behavioral sciences---is a Bayesian approach used to assess the Hurst-Kolmogorov (HK) process in hydrology \cite{tyralis2014bayesian}. In this method---which we call the "HK method," Tyralis and Koutsoyiannis \cite{tyralis2014bayesian} proposed a Bayesian-inspired technique that defines the posterior distribution from which to sample $H$.

\vspace{6 pt}

We previously compared the performance of the HK method and the DFA using simulated and empirical time series \cite{likens2023better}. Using synthetic time series with \textit{a priori} known values of $H$, we demonstrated that the HK method consistently outperforms DFA in three ways. The HK method (i) accurately assesses long-range correlations when the measurement time series is short, (ii) shows minimal dispersion about the central tendency, and (iii) yields a point estimate that does not depend on the length of the measurement time series or its underlying Hurst exponent. Furthermore, comparing the two methods using empirical human behavioral time series supported these simulation results. We also showed that the HK method balances the Type I and Type II errors associated with inferential statistics performed on the estimated $\hat{H}$ (We use $\hat{H}$ to distinguish the estimated value of the Hurst exponent from the ground truth $H$). It reduces the likelihood of the Type II error by not missing an effect of an independent factor when it exists, without increasing the likelihood of the Type I error by finding an effect of an independent factor when it does not exist. DFA nonetheless confers an advantage in computing time---owing to the simple and linear nature of computations, even though these results provide a convincing argument for choosing the computationally-expensive HK method over DFA. Therefore, computational efficiency, particularly for high throughput and real-time applications, is critical to successfully implementing the HK method.

\vspace{6 pt}

The HK method is computationally expensive, owing to its roots in the Bayesian framework. The computation time increases exponentially with the time series length $N$. When performed on a personal computer, the computation time could easily exceed several hours for $N \geq 1024$---typical time series length in behavioral sciences (e.g., stride interval time series, RT time series). This problem becomes even more challenging when dealing with physiological measurements recorded over longer times (e.g., breathing rate variability, heart rate variability, functional near-infrared spectroscopy, fNIRS) or at higher frequencies (e.g., the center of pressure, CoP, electroencephalogram, EEG, electromyography, EMG). Moreover, the computational limitation makes it impractical to implement the HK method in real-time paradigms, such as biofeedback and brain-computer interfaces. Computationally optimizing the HK method for accurately estimating $H$, i.e., $\hat{H}$, is, therefore, critical for promoting the adoption of the HK method as a standard approach to  estimating the Hurst exponent in behavioral sciences.

\vspace{6 pt}

Here, we provide data on the accuracy of the Hurst exponent, estimated using the HK method for synthetic time series as a function of \textit{a priori} known values of ${H}$, time series length, and the number of samples from the posterior distribution of $H$---a parameter related to the Bayesian estimation that critically influences the accuracy of $\hat{H}$. Our results will guide the selection of the minimum sample of $H$ from the posterior distribution of $H$ necessary for estimating $\hat{H}$ for a given level of accuracy.

\section*{Methods}

\subsection*{The HK method for estimating the Hurst exponent}

As noted above, a recently introduced Bayesian approach to estimating $H$ \cite{tyralis2014bayesian} shows remarkable promise in addressing fundamental limitations with DFA. In previous work, we have demonstrated that the HK method outperforms DFA in several contexts \cite{likens2023better}. Our current interest is investigating the HK method’s performance trade-offs related to computational efficiency. Results presented later in Section \ref{results} demonstrate that the HK method is entirely accurate in recovering $H$ from time series even when sacrificing some accuracy for enhanced computational efficiency. Below, we provide a brief overview of the HK method while referring the reader to the foundational work for additional mathematical details and proofs \cite{tyralis2014bayesian}. Our notation generally follows that original work.

\vspace{6 pt}

The foundation for the method originates in the definition of the autocorrelation function for the so-called Hurst-Kolmogorov (HK) process \cite{koutsoyiannis2003climate} as:

\begin{align*}
\rho_{k} = |k + 1|^{2H}/2 -2|k|^{2H}/2 + |k - 1|^{2H}, \quad k = 0,1,\dots, \tag{1}\label{eq:1}
\end{align*}

\noindent{}such that $H$ is the Hurst exponent, $k$ is the time lag, and $\rho_{k}$ is the autocorrelation function at each successive value of $k$. If $H = 0.5$, then $\rho_{k}$ is 1 when $k=0$ but zero for $k > 0$. If $0 < H < 0.5$, then $\rho_{k}$ is negative at lag $1$ before damping zero when $k > 1$. Lastly, if $0.5 < H < 1$, then $\rho_{k}$ is positive and slowly decays towards zero; and as $H \rightarrow 1$, $\rho_{k}$ asymptotically approaches $0$.

\vspace{6 pt}

As noted, the HK method is a Bayesian approach to estimating $H$ \cite{tyralis2014bayesian}. In the foundational work, Tyralis \& Koutsoyiannis \cite{tyralis2014bayesian} derived a method to sample from the posterior distribution of $H$ given by:

\begin{align*}
\pi(\boldsymbol{\varphi}|\textbf{x}_{n}) \propto |\textbf{R}_{n}|^{-1/2} \: [\textbf{e}_{n}^{T} \textbf{R}_{n}^{-1} \textbf{e}_{n} \textbf{x}_{n}^{T} \textbf{R}_{n}^{-1} \textbf{x}_{n} - (\textbf{e}_{n}^{T} \textbf{R}_{n}^{-1} \textbf{e}_{n})^{2}]^{-(n-1)/2}\\ (\textbf{e}_{n}^{T} \textbf{R}_{n}^{-1} \textbf{e}_{n})^{n/2 - 1}, \tag{2}\label{eq:2}
\end{align*}

\noindent{}The natural logarithm of Eq. (\ref{eq:2}) is then given by:

\begin{align*}
\ln{\pi(\boldsymbol{\varphi}|\textbf{x}_{n})} \propto \frac{1}{2} \ln{|\textbf{R}_{n}|} \: -\frac{(n-1)}{2} \ln{[\textbf{e}_{n}^{T} \textbf{R}_{n}^{-1} \textbf{e}_{n} x_{n}^{T} \textbf{R}_{n}^{-1} \textbf{x}_{n} - (\textbf{e}_{n}^{T} \textbf{R}_{n}^{-1} \textbf{e}_{n})^{2}]}\\ +
\frac{n - 2}{2} \ln{(\textbf{e}_{n}^{T} \textbf{R}_{n}^{-1} \textbf{e}_{n})}, \tag{3}\label{eq:3}
\end{align*}

\noindent{}where $\textbf{R}_{n}$ is the autocorrelation matrix with elements $r_{i,j}$ where $i, j = 1, 2, 3, \dots, n$, $\textbf{e}_{n} = (1, 1, 1, \dots, 1)^{T}$ is a vector of ones with $n$ elements, $|\dots|$ notes a determinant, the superscript of $-1$ in $\textbf{R}_{n}^{-1}$ is a matrix inverse, and the superscript $T$ is a matrix transpose. The right-hand products in Eq. (\ref{eq:3}) are derived from the quadratic forms of the inverse of a symmetric, positive definite autocorrelation matrix (Levinson Algorithm; Algorithm 4.7.2, Golub \& Van Loan \cite{golub2013matrix}, p. 235) for a given $x_{t}$ and $\rho_{k}$.

\vspace{6 pt}

Accept-reject algorithms are standard tools for sampling from posterior distributions and serve as the backbone of implementing the HK method \cite{robert1999monte}. Let $f(x)$ be a probability density function (PDF) from which it is difficult to sample. $f(x)$ is the ``target distribution" and can be sampled using Monte Carlo methods. First, one samples a simpler ``proposal distribution," $Mg(x)$ that has the same domain as $f(x)$ and $M$ is a constant large enough to ensure that $g(x) \geq f(x)$. In theory, the proposal distribution, $g(x)$, can be any number of distributions, such as uniform, Gaussian, exponential, etc. However, the algorithm gains computational efficiency when the overall shape of $g(x)$ is similar to $f$. Second, $f(x)$ is evaluated at the value obtained by sampling $g(x)$, the proposal distribution. Third, a sample is drawn from $U(x) \sim Uniform(0, Mg(x))$. If $U(x) \leq f(x)$, then the value proposed by sampling $g(x)$ is accepted as a valid sample. Otherwise, the proposal is rejected, and the algorithm is re-initialized. This process repeats until $n$ samples are obtained, where $n$ is the number of samples desired from the posterior distribution. We seek to understand appropriate values of $n$ that balance estimation accuracy and computational efficiency.

\vspace{6 pt}

In experiments reported in Section \ref{experimental_design} and Section \ref{results}, we employ this accept-reject algorithm to sample $H$ from its posterior distribution (Algorithm A.5, Robert \& Casella \cite{robert1999monte}, p. 49). The target distribution, $f(x)$ is Eq. (\ref{eq:3}) and $g(x) \sim Uniform(0,1)$. That uniform distribution makes sense for $g(x)$ because it shares the same domain of $H$ and hence Eq. (\ref{eq:3}), namely $(0,1)$ \cite{tyralis2014bayesian}. A numerical optimization routine is used to determine $M$ by finding the maximum of Eq. (\ref{eq:3}) as a function of $H$. The point estimate of $H$ is then taken as the median of the posterior distribution of $H$. Time series were analyzed using the $R$ \cite{team2013r} programming environment using the \texttt{inferH()} function as part of the ``HKprocess" package \cite{tyralis2022hkpackage}.

\subsection*{Generating synthetic time series with \textit{a priori} known values of the Hurst exponent}\label{experimental_design}

The Davies-Harte algorithm \cite{davies1987tests} was used to  generate fractional Gaussian noise (fGn), which can be tuned to exhibit varying degrees and direction of autocorrelation consistent with Eq. (\ref{eq:1}).  fGn time series were generated in \texttt{R} \cite{team2013r} using the function \texttt{fgn\_sim()} from the package ``fractalRegression" \cite{likens2021fractalregression}. The function \texttt{fgn\_sim()} has two inputs: the time series length, $N$, and the Hurst exponent, $H$. We generated $1,000$ synthetic fGn time series for each combination of six different time series lengths ($N = 32, 64, 128, 256, 512, 1024$) and nine different \textit{a priori} known values of $H$ ($H = 0.1, 0.2, \dots, 0.9$). We submitted all synthetic time series to the HK method in \texttt{R} \cite{team2013r} using the function \texttt{inferH()} from the package ``HKprocess" \cite{tyralis2022hkpackage}. The function \texttt{inferH()} has two inputs: the time series, $x_{N}$, and the simulated sample size from the posterior distribution of $H$, $n$. The HK method was performed for a progressively larger sample from the posterior distribution of $H$, covering an extensive range with the sample size ranging from $1$ to $25$ with increments of $1$ and from $25$ to $500$ with increments of $25$, i.e., $n = {1, 2, \dots, 25, 50, \dots, 500}$.

\section*{Results} \label{results}

\begin{figure*}
\includegraphics[width=4.5in]{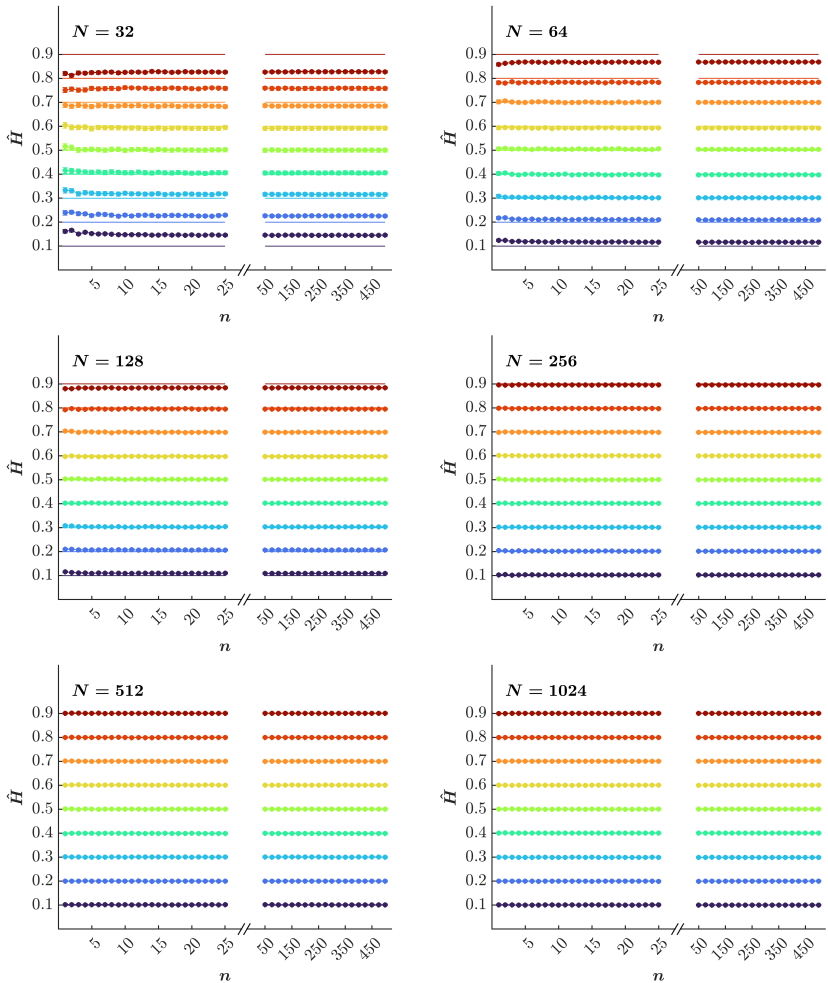}
\caption{\textbf{The Hurst exponent, \boldmath$\hat{H}$, estimated using the HK method closely matches the actual $H$ for time series containing as few as $128$ values.} Each panel plots the $Mean$ $H$ for $1,000$ synthetic time series with $N = 32, 64, 128, 256, 512, 1024$, \textit{a priori} known values of $H$ ranging from $0.1$ to $0.9$, and the sample size of $H$ desired from the posterior distribution of $H$, $n = {1, 2, \dots, 25, 50, \dots, 500}$. Horizontal colored lines indicate the actual $H$, and error bars indicate $95$\% CI across $1000$ simulations.}
\label{fig:HKprocess_1}
\end{figure*}

\begin{figure*}
\includegraphics[width=4.5in]{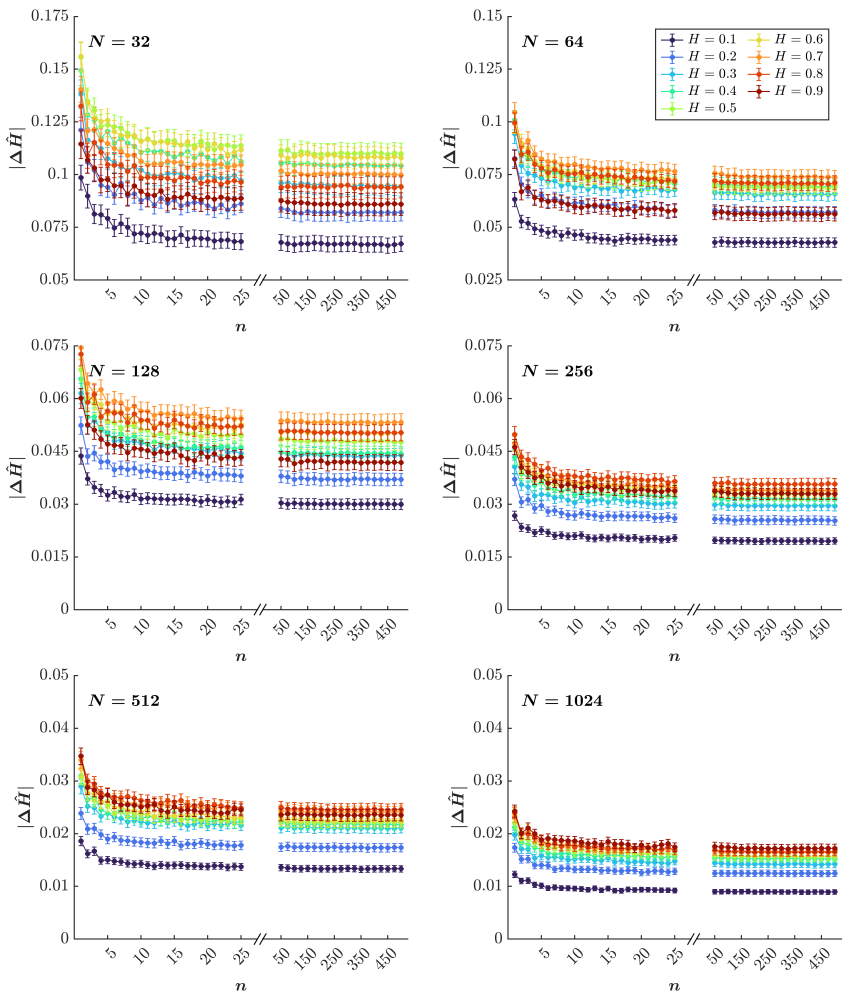}
\caption{\textbf{The absolute error in the estimation of the Hurst exponent, \boldmath$\hat{H}$, using the HK method sharply reduced with the sample size of \boldmath$H$ desired from the posterior distribution of  \boldmath$H$, reaching an asymptote at \boldmath$n = 50$.} Using a larger simulated sample from the posterior distribution of $H$---i.e., $n > 50$---does not influence the accuracy of the estimated $H$. Each panel plots the $Mean$ absolute error in the estimated $H$ for $1,000$ synthetic time series with $N = 32, 64, 128, 256, 512, 1024$, \textit{a priori} known values of $H$ ranging from $0.1$ to $0.9$, and the sample size of $H$ desired from the posterior distribution of $H$, $n = {1, 2, \dots, 25, 50, \dots, 500}$. Error bars indicate $95$\% CI across $1000$ simulations.}
\label{fig:HKprocess_2}
\end{figure*}

Figs. \ref{fig:HKprocess_1} \& \ref{fig:HKprocess_2} provide a summary visualization of the simulation results for each combination of the \textit{a priori} known values of the Hurst exponent ($H = 0.1, 0.2, \dots, 0.9$), the time series length ($N = 32, 64, \dots, 1024$), and the sample size of $H$ desired from the posterior distribution of $H$, $n = {1, 2, \dots, 25, 50, \dots, 500}$. As a general preview, $\hat{H}$ estimated using the HK method closely matches the actual $H$ for time series containing as few as $128$ values (Fig. \ref{fig:HKprocess_1}, middle left). For shorter time series---$N = 32, 64$, the HK method visibily overestimates $\hat{H}$ for smaller actual $H$ and underestimates $\hat{H}$ for larger actual $H$ (Figs. \ref{fig:HKprocess_1}, top left and top right, respectively). Shorter time series---$N = 32, 64, 128$---show more closely matching values of $\hat{H}$ and $H$ for larger posterior samples of $H$. Still, this trend was not apparent for longer time series---$N = 256, 512, 1024$ (Fig. \ref{fig:HKprocess_1}, middle right, bottom left, and bottom right, respectively). Nonetheless, the sample size of $H$ desired from the posterior distribution does not seem to influence the fidelity of $\hat{H}$ for a sufficiently large sample size of $H$ desired from the posterior distribution, i.e., $n = 50$.

\vspace{6 pt}

When $N = 32$, a very short time series compared to the DFA standard of $> 500$, the estimated $\hat{H}$ shows a large discrepancy ($> 0.1$) with the actual $H$ in terms of the absolute error (Fig. \ref{fig:HKprocess_2}, top left). Although this discrepancy sharply reduced with the sample size of $H$ desired from the posterior distribution, almost reaching an asymptote by $n = 25$, the absolute error remains considerably high even for $n = 500$. For a relatively longer yet considerably short time series---$N = 64$, the absolute error reduces with the sample size of $H$ desired from the posterior distribution, reaching a lower asymptotic value of $< 0.1$ (Fig. \ref{fig:HKprocess_2}, top right). However, this discrepancy is still sufficient to mask the typically observed differences in the Hurst exponent of empirical time series across two groups. For instance, the Hurst exponent of stride-to-stride interval time series during walking shows a difference of the order of 0.1 across various task conditions(e.g., \cite{mangalam2022leveraging,raffalt2021temporal,raffalt2023stride}. However, a discrepancy of the order of 0.1 in the estimation of $H$ can potentially mask these differences, resulting in false negatives in statistical tests. The trend was comparable for time series with $N = 124$ except for the absolute error reaching a lower asymptotic value of $\sim 0.05$ (Fig. \ref{fig:HKprocess_2}, middle left).

\vspace{6 pt}

When $N = 256$, the absolute error in the estimated $\hat{H}$ falls within a much lower range of tolerance ($< 0.05$) even with a very small sample of $H$ from the posterior distribution (Fig. \ref{fig:HKprocess_2}, middle right). Again, the absolute error sharply reduces with $n$, reaching an even lower asymptotic value of $\sim 0.04$ by $n = 25$. Longer time series with $N = 512$ and $N = 1024$ also show similar trends except for much lower asymptotic values of absolute error: $< 0.03$ and $< 0.02$, respectively, by $n = 25$ (Fig. \ref{fig:HKprocess_2}, bottom left and bottom right, respectively). In short, a relatively small sample size of $H$ desired from the posterior distribution when using the accept-reject algorithm of the HKM suffices to estimate $\hat{H}$ with very high accuracy. Increasing the sample size of $H$ desired from the posterior distribution considerably increases the computational cost but confers no additional advantage in terms of the accuracy of estimating the Hurst exponent.

\section*{Discussion}

The HK method offers several advantages over DFA in estimating the Hurst exponent of a time series \cite{likens2023better}. However, these advantages come at the cost of computation time. The HK method is computationally expensive owing to its roots in the Bayesian framework. Computationally optimizing the HK method for accurately estimating $\hat{H}$ is, therefore, critical for promoting the adoption of the HK method over DFA for estimating the Hurst exponent in behavioral sciences. To address this issue, we have provided data on the accuracy of the Hurst exponent estimated using the HK method for synthetic time series as a function of \textit{a priori} known values of ${H}$, the time series length, and the sample size of $H$ from the posterior distribution of $H$---a parameter related to the Bayesian estimation that critically influences the accuracy of $\hat{H}$ and computation time. The simulated sample from the posterior distribution of $H$ as small as $n = 50$ suffices to estimate the Hurst exponent with reasonable accuracy. Using a larger simulated sample from the posterior distribution of $H$---i.e., $n > 50$---provides only a marginal gain in accuracy, which might not be worth trading off with computational efficiency. We suggest balancing the simulated sample size from the posterior distribution of $H$ with the computational resources available to the user, preferring a minimum of $n = 50$ and opting for larger sample sizes based on time and resource constraints. Our results allow the reader to make such judgments.

\vspace{6 pt}

Empirical data pose several other issues that might undermine the accuracy of $H$ estimated using the HK method, such as strong trends \cite{bryce2012revisiting,hu2001effect,horvatic2011detrended}, nonstationarities \cite{bryce2012revisiting,chen2002effect}, and ``crossovers"---i.e., when correlations do not follow the same scaling law for all timescales and a crossover is observed between different scaling regions \cite{bashan2008comparison,kantelhardt2001detecting,kelty2013tutorial,peng1995quantification}. However, because these anomalies, particularly the crossovers, often appear at longer timescales, it seems reasonable to promote the systematic use of this method, irrespective of these reserves. Future work could investigate how trends, nonstationarities, and crossovers influence the estimation accuracy using the Bayesian approach.

\vspace{6 pt}

In summary, the HK method offers several advantages over DFA for estimating the Hurst exponent of a time series, especially when it is short. What may prevent, however, the adoption of the HK method is significantly long computation time, especially when analyzing time series containing several hundred to thousands of measurements---which is frequently the case in behavioral sciences. To minimize the computation time of the HK method, we have provided critical information for identifying the minimum simulated sample from the posterior distribution of $H$. This information could aid the selection of correct parameters that allow the estimation of $H$ in real-time settings, such as biofeedback paradigms and brain-computer interfaces, where computation time is often limited.

\backmatter

\bmhead{Acknowledgments}
This work was supported by the Center for Research in Human Movement Variability at the University of Nebraska at Omaha, the University of Nebraska Collaboration Initiative, the NSF award 212491, the NIH awards P20GM109090 and R01NS114282, the NASA EPSCoR mechanism, and the IARPA WatchID award.

\bmhead{Author contributions}
Conceptualization: M.M. and A.D.L.; Methodology: M.M., T.W., J.H.S., and A.D.L.; Formal analysis: M.M.; Data curation: M.M.; Writing -- Original draft: M.M.; Writing -- Review \& Editing: M.M., T.W., J.H.S., and A.D.L.; Visualization: M.M.; Funding acquisition: A.D.L.

\bmhead{Declarations}
The authors declare no competing financial interests.

\bibliography{sn-bibliography}

\end{document}